# Hybrid Social Networking Application for a University Community

**O. Chigozie**

Department of Computer Science
University of Port Harcourt
Rivers State, Nigeria
*E-mail: ogueri@gmail.com*
*Tel: +234 7*

**P. Williams**

Department of Computer Science
University of Calabar
Rivers State, Nigeria
*E-mail:williams@gmail.com*
*Tel: +234 8*

**N.E. Osegi**

Department of Information and Communication Technology
National Open University of Nigeria
Lagos State, Nigeria
*E-mail: geeqwam@gmail.com*
*Website: www.osegi.com*
*Tel: +234 7030081615*

**ABSTRACT**

A hybrid social network for building social communities for a university community is presented. The system employed the semantic ontology for an offline/online social network site (SNS) using a Mobile Ad Hoc Network. It captures the core features of an SNS including profile creation, friend invite/search, group formation, chatting/messaging, blogging and voting. Three core frameworks – the peer2me framework, SMSN semantic mobile social network framework and Peoplepods framework were considered in the implementation phase. The results show remarkable matching performance for prosumers with similar interests with relevance close to unity. The social network was able to capture the needs of the university students by serving as a handy direction to popular locations within the campus.

## 1. INTRODUCTION

A social networking service is an online service, platform, or site that focuses on facilitating the building of social networks or social relations among people (or organization or other social entity) who, for example, share interests, activities, backgrounds, or real-life connections. Online social networks are defined as Web-based services that allow individuals to create a profile, make a list of users with whom they share a connection, and review their list of contacts [ (Boyd & Ellison, 2007)]. These individuals may also be regarded as prosumers [2] – a generic social networking term used to refer to one who generates and consumes social content.

Today mobile phones are often equipped with several network technologies such as the General System for mobile Telecommunication (GSM), Universal Mobile Telecommunications System (UMTS), Bluetooth and Wireless Local Area Network (WLAN). These technologies make it possible to connect mobile phones to other devices in several ways all of which have different advantages and disadvantages. Several mobile phones that interconnect using a network technology such as Bluetooth make up a Mobile Ad Hoc Network (MANET). The vast functionality of mobile phones combined with state of the art network technology makes them ideal as mobile collaborative devices.

The fundamental objectives for mobile social networks are adaptability, flexibility, cost effectiveness and easy deployment. Presently social networking links are done mostly through the internet with a few of them operated via intranet and other short range communication links such as blue tooth and WiFi. Hence being able to adapt to the various transmission links would make for a more affordable, dependable, and efficient communication system.

## 2. PROBLEM STATEMENT

The need to share information, interact with others, and be fully informed about ones environment have become very important and challenging today, especially within the academic environment.

Presently many social networking applications tend to not only focus on interaction with peers, and offering entertainment but also depend solely on ether online or offline communication links. This has been a very serious problem since in the event of an online network failure; there is a total brake down in communication. Even with the offline communication technology, users are still limited either in their inability to store ever increasing data on their memory card (SDCARD) or their inability to connect with peers due to the range in

To make information readily accessible to people, especially within a specific social community, there is a need for a robust system that is more efficient, affordable and secure.

This low cost of information access and dissemination and also the bridging of the gap between online and offline communication is what this work seeks to address.

frequency signals.To solve this problem a hybrid social network application code named MeYounMe(MYM) shall be developed. This work is a slight shift from the existing social networking archetype towards a mobile ad hoc network (MANET) that potentially combines the technique to connects all types of devices that are equipped with short-range communication medium, such as Bluetooth, WiFi, and the local intranet and internet. The current version also requires a great deal of time to comprehend and may only currently provide communication via Bluetooth's alone; although it is also intended to implement the WiFi technology. It may therefore be necessary to redesign the architecture to incorporate the WiFi in order to make it easier for other developers to utilize the framework.

## 3. SNS METHODOLOGY

In this section, a hybrid methodology incorporating three frameworks, peer2me, SMSN and Peoplepods will be presented.

### 3.1 Social Graphs

A graph is a mathematical abstraction for modelling relationships between things. A graph is constructed from nodes (the things) and edges (the relationships). This mathematical tool that can model natural and artificial systems such as economy, diseases, power grids, etc. has been used by the anthropologists, sociologists and other humanities oriented academics. However, graph analysis and social network analysis are also valuable tools for studying the web and human behaviours of the web users [2].

Social network analysis may be applied in any web field where a graph may be constructed. From the appearance of social networking sites, users were forming graphs with their friends and this was the ideal source of fresh data to apply social network analysis. One of the most prominent issues in social networks is the formation or the identification of a network of nodes based on real world knowledge (school friends, colleagues etc.) or web extracted knowledge (they are part of the same online community, they like the same movies, etc.).

*Social graph expansion*

Social network analysis is applied in the web by utilising the interconnected Web 2.0 blogs and their comments.

Backlinks of posts and the blogroll (list of other blogs) of each blog constructed a graph that could provide some information. This structure was difficult to update, error prone (copy paste links, write urls, etc.) and the users had to have a web page or blog of their own.

Social networking sites created the tool that made relations easier to track and build. Now every user that has an account in a SNS can "tag" information and propagate it to that network. "Likes", "tweets", "diggs", etc. are one button actions that users perform while surfing the web in order to post a piece of information without leaving from the current web page.

*Facebook open graph*

Facebook Open Graph provides an interactive interface for interconnecting web pages/content with the Facebook social graph links. The most common practice is to add a "Like" button near a media object such as a video or music file in a web page and let prosumers share their "Likes". When a prosumer clicks the "Like" button outside the Face-book platform, a new connection is formed in the prosumer's profile and the prosumers can have access to the file. This simple API has significant impact on the generated content in the Facebook platform considering the over 500 million active Facebook users that surf the web and collect "Likes".

**3.2 Peer2Me Framework**

Peer2Me is a framework that enables source code developers to create collaborative applications for mobile phones using a network technology such as Bluetooth. It enables developers to focus on the logic of the actual application and takes care of management data exchange and network configuration in the P2P network [3]. It is based on the Java 2 Micro Edition programming and development platform. Peers can exist with variable identities that possess a name description of a particular function. The architecture of a Peer2me framework is shown in Fig1.

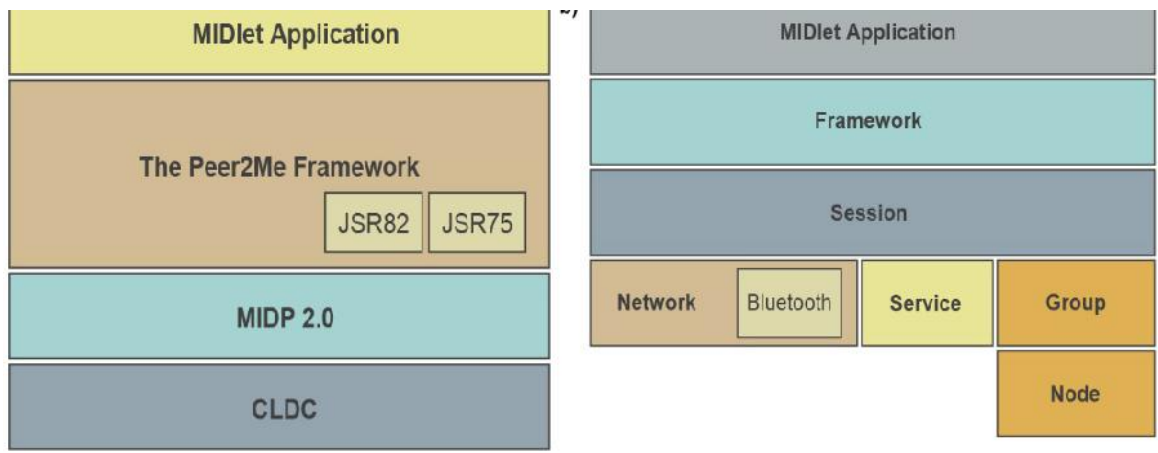

Figure1. Peer2me Architecture (Courtesy, Wang and others)

**3.3 SMSN Framework**

SMSN is a framework for building semantically structured adhoc peer to peer SNSs for mobile phones using a network technology such as Bluetooth. In SMSN, a so called T-Box ontology defines the common understanding for all the important concepts and their relationships [4] . For the creation of the T-Box ontology a top-down approach is adopted; this starts with important general concepts such as Person, Name, Age, and Gender, which can later be enriched and specialized. Dynamic concepts such as current activity, orientation, motion state, current terminal and location are also featured. From the architecture, profiles can be created and stored by the user. Based on relevance criterion and for a 2-person network a similarity match is activated during a friend search if their profiles match. This concept of Profile similarity has been defined in [4]. The similarity concept is a valuable criterion for measuring the probability of a match between two profiles in the network. In T-Box ontology, concepts and profiles may be defined in terms of their distances and similarity from and between one another respectively. In the T-Box the profile summary, profile similarity, concept distance, and concept similarity are featured. By introducing the concept of relevance, profile similarity measure can be enforced by the program. Thus, profiles with similar interests should possess a high relevance while those with dissimilar interests a low relevance. Thus, relevance may be defined in this context as the probability that two interests match. Extending our definition further, one may define relevance as the probability of two or more similar interests occurring in a T-Box.

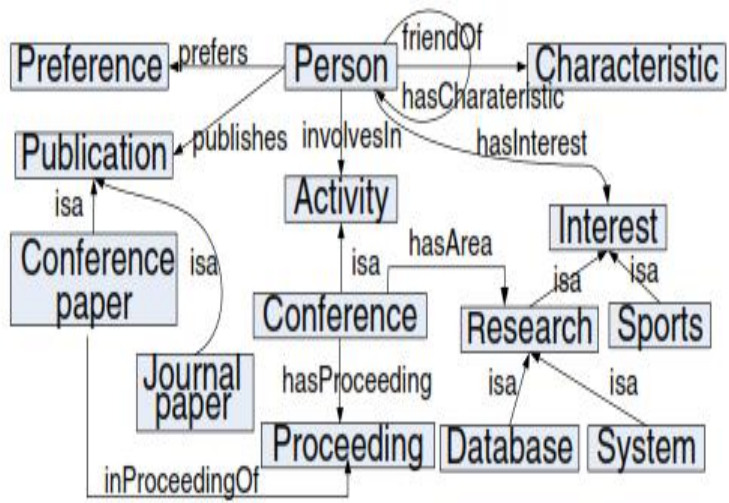

Figure2. Part of a T-Box ontology (Coutesy, Li and others, 2009)

### 3.4 Peoplepods Framework
**PeoplePods**

PeoplePods is an object-oriented web-based framework that makes it easier to create community or "social" applications [5]. It provides a flexible infrastructure within which members of a site can create, comment upon, and consume content of arbitrary types. Popular social functionality such as user profile creation, friend lists; personalized content views, bookmarking, and voting are automatically available within any PeoplePods application. PeoplePods can be used to build stand-alone sites, or can be layered onto existing sites such as those run using Wordpress or other PHP applications. The Peoplepods architecture featuring some core SNS features is as shown in Fig 3. Once a user profile is created a person object is incarnated into the SN and the person can then perform the basic functions in an SNS. The site administrator typically moderates the functions of the "basic prosumer" by carrying out super functions within the site keeping a watchful eye on the events that occur. The site administrator is the "Super Prosumer".

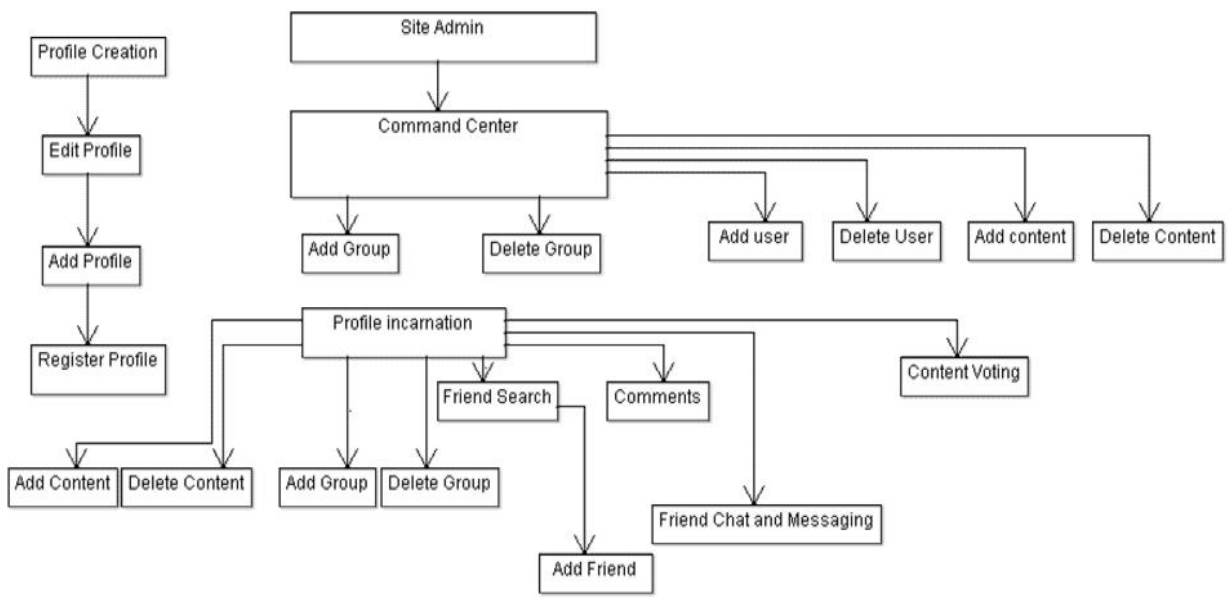

Figure3. The Peoplepods Architecture

## 4. RESULTS

The results of the tests are presented in this section. The results are run for the offline (chat2me and MobisnMIDlet) applications and the online (web2me.jar) applications. An offline emulation study has been implemented using the Netbeans 7.1.2 software tool. Two users are emulated to achieve a minimalist network. The emulation capture for the chat2me application and MobisnMidlet applications are as shown in Fig 4, 5 and 6 respectively. In Fig 4 the chat process is initiated by the incarnation of two chat2me MIDlets using the Netbeans default platform emulator. The process of emulation takes the form of a person –to-person interaction wherein common interests may be found. The same is also true for the offline SN but this time more parameters are involved. Fig 5 captures a friend search while Fig 6 shows the relevance (approximately 0.99999) in a prosumers (Nkechi) when a match is discovered. Access to the online network (not shown) is gained using a technique referred to as Platform request.

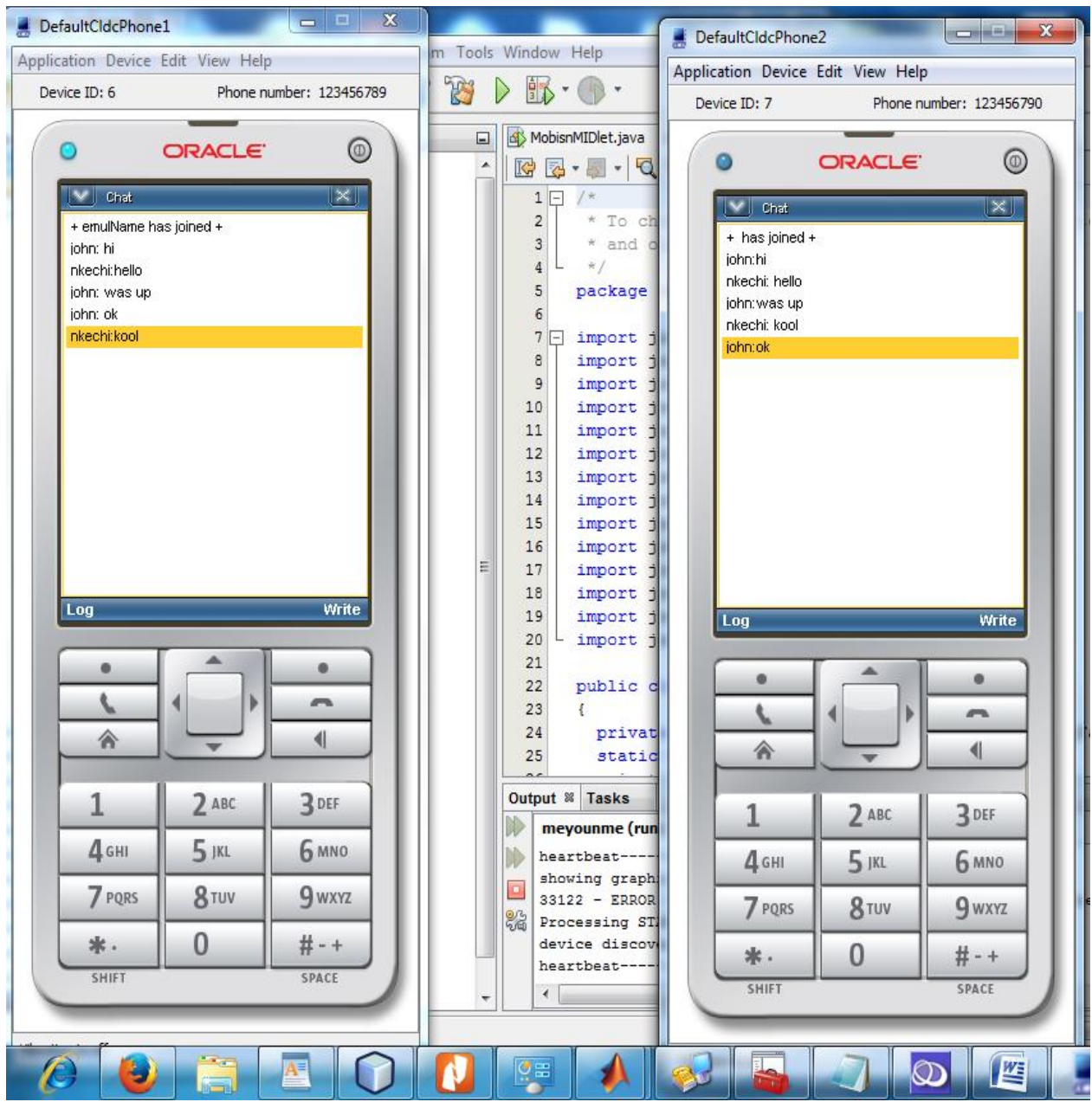

Figure4. Chatting between two peers: Nkechi and John

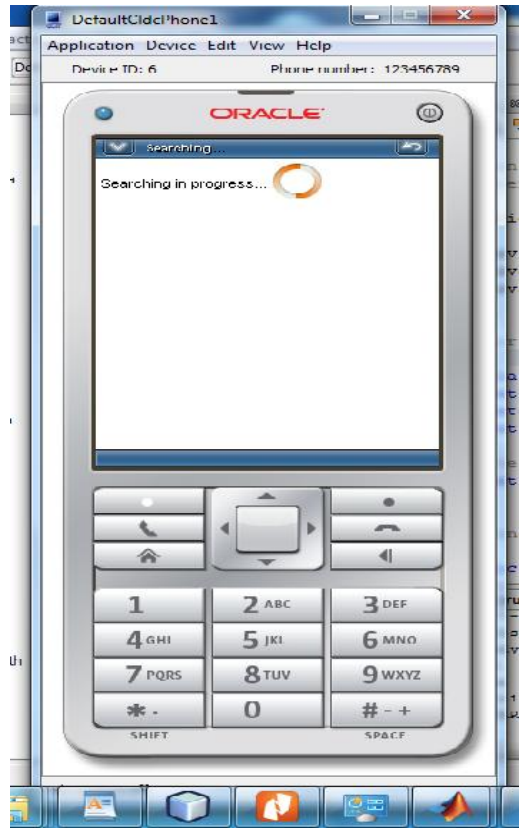

Figure5. A Friend search

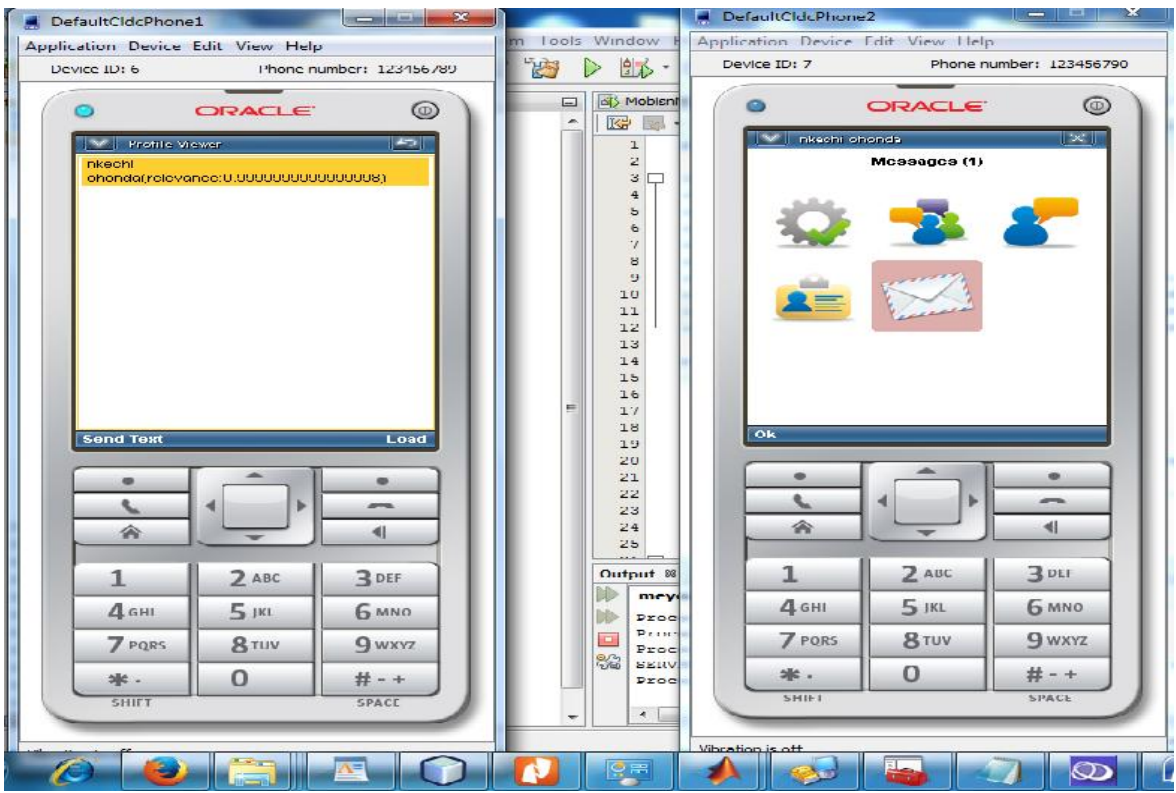

Fig 6. Snapshot showing a relevance of 0.9999 for user 1

## 5. DISCUSSIONS

From the theory of similarity and the concept of social networking graphs, three frameworks have been examined and applied in the context of a social network. The social network MeYounMe is able to implement the core features of SNS, friend matchmaking, profile creation, blogging, and group formation. This has been applied to to a university community and is tested with user content.

The following postulates can be made based on the observed social network developed:

(i) For a social graph to be formed, there is need for at least two prosumers in the interaction process
(ii) In a SNS, a prosumer is an incarnation of a person
(iii) A prosumer can synthesise social data, form groups, make a friend or content search, create blogs (messaging arms) and administer a site.
(iv) The relevance of a monitoring prosumer may be determined from the interests compared to a reference prosumer for a given instant of time.
(v) If the relevance tends to unity, then the prosumers match otherwise they differ
(vi) The margin by which a relevance point might differ is dependent on the viewpoint from the observer. If the observer is near then relevance might increase, if the observer is far the converse is possible.

W

## 6. CONCLUSIONS

A hybrid SNS that features all the core principles of SNS and mobile adhoc dcentralized SN systems have been developed and tested for functionality including Profile creation, friend search, messaging and profile match-making and tagging using the Bluetooth and internet communication technologies. An off-line chat mechanism has been captured using the peer2me framework in j2me compliant mobile device. It has been shown using semantic reasoning based on similarity constructs and the concept of relevance; how an adhoc decentralized SN might freely interact.

## 7. RECOMMENDATIONS FOR FUTURE WORK

Social network applications that tend to be platform independent are gaining more attention. Thus, there is need to develop content in this area to ensure continuous support even when the system is accessed from different terminals. Some aspects of the SN system e.g. content rendering may still need to be adapted to suit various phone models. There is also the need to address the issue of security and spam infiltration into the SNS. Spam infiltration can be a nuisance particularly for small useful sites and can lead to network turbulence. We recommend a platform independent, spam resistant and secure Social Network System for further investigations.